\documentclass{ws-ijmpa}

\begin{document}

\markboth{Marcelo J. Rebou\c{c}as}
{Constraints on dark energy and Cosmic Topology}

%
\catchline{}{}{}{}{}
%

\title{CONSTRAINTS ON DARK ENERGY AND COSMIC TOPOLOGY}

\author{MARCELO J. REBOU\c{C}AS}

\address{Centro Brasileiro de Pesquisas F\'{\i}sicas
\\ Rua Dr. Xavier Sigaud 150, \ 
22290-180 \ Rio de Janeiro -- RJ, Brazil \\
reboucas@cbpf.br} 

\maketitle

\begin{history}
\received{Day Month Year}
\revised{Day Month Year}
\end{history}

\begin{abstract}
A non-trivial spatial topology of the Universe is a potentially  
observable attribute, which can be probed through the circles-in-the-sky 
for all locally homogeneous and isotropic universes with no assumptions 
on the cosmological parameters.
We show how one can use a possible circles-in-the-sky detection 
of the spatial topology of globally homogeneous universes to set 
constraints on the dark energy equation of state parameters.
\end{abstract}

\ccode{PACS numbers: 98.80.Es,98.80.-k,  98.80.Jk}

\section{Introduction}  \label{Intro}

In the context of general relativity, the observable Universe
seems to be well described by a $4$-manifold
$\mathcal{M} =\mathbb{R}\times M$ with locally homogeneous and
isotropic spatial sections $M$, and therefore endowed with a
Robertson--Walker metric
\begin{equation} \label{RWmetric}
ds^{2}=-dt^{2}+a^{2}(t)\left[ d\chi^{2}+S_k^{2}(\chi)(d\theta
^{2}+\sin^{2}\theta d\phi^{2}) \right]  \,, 
\end{equation}
where $a(t)$ is the scale factor and $S_k(\chi)= \chi\,,\sin\chi\,,\sinh\chi\,$ 
depending upon whether the geometry of the spatial sections is Euclidean, 
spherical or hyperbolic with constant spatial curvature $k=0$, $1$, or $-1$.
The spatial geometry or the corresponding spatial curvature is an observable property, 
which can be determined by finding out whether the total energy-matter density 
of the Universe, $\Omega_{\mathrm{tot}}$, is equal to, greater than or small 
than 1.
In consequence, a key point in the search for the spatial geometry of
the Universe is to use observations to constrain the density
$\Omega_{\mathrm{tot}}$.
Often the homogeneous and isotropic spatial sections $M$ are
assumed to be the simply connected $3$-manifolds: Euclidean $\mathbb{R}^{3}$,
spherical $\mathbb{S}^{3}$, or hyperbolic $\mathbb{H}^{3}$.
However, the $3$-space $M$ can also be one of the possible
quotient (multiply-connected) manifolds $\mathbb{R}^3/\,\Gamma$, 
$\mathbb{S}^3/\,\Gamma$, and $\mathbb{H}^3/\,\Gamma$, where $\Gamma$ is 
a fixed-point free discrete group of isometries of the corresponding 
covering space $\mathbb{E}^3$, $\mathbb{S}^3$, or $\mathbb{H}^3$.
The local geometry of the spatial sections $M$ thus constrains,
but does not dictate, its topology (see, e.g., the 
review Refs.~\refcite{CosmTopReviews}). 

The immediate observational consequence of a detectable
multiply-connected spatial section $M$ is that an observer 
could potentially detect multiple images of radiating sources. 
In this way, in a universe with a detectable\cite{TopDetec} non-trivial 
topology the  last scattering surface (LSS) intersects some of its 
topological images in the so called circles-in-the-sky,\cite{CSS1998} 
i.e., pairs of matching circles of equal radii, centered at different 
points of the LSS with the same distribution of temperature fluctuations
(up to a phase) along the circles of each pair.
Therefore, to observationally probe  a non-trivial spatial topology 
on the largest available scales, one needs to scrutinize the cosmic 
microwave background (CMB) sky-maps in order to extract such correlated 
circles, and use their angular radii, the relative phase and position 
of their centers to determine the spatial topology of the Universe.
Hence, a detectable non-trivial cosmic topology is an observable attribute, 
which can be probed through the circles-in-the-sky for all locally
homogeneous and isotropic universes with no assumptions on the
cosmological parameters.

The question as to whether one can use the knowledge of the topology 
to either determine the geometry or set constraints on the density 
parameters naturally arises here. Regarding the geometry it is well-known 
that the topology of $M$ determines the sign of its curvature (see, e.g.,
Ref.~\refcite{BernshteinShvartsman1980}) and therefore the $3$-geometry. 
Thus, the topology of the spatial section of the Universe dictates its geometry.
In recent works,\cite{Previous1}\cdash\cite{BBRS2006b} it has been shown that 
the knowledge of a \emph{specific} spatial topology through the circles-in-the-sky 
offers an effective way of setting constraints on the density parameters 
associated with matter ($\Omega_m$) and dark energy ($\Omega_{\Lambda}$) 
in the context of $\Lambda$CDM model. 
In other words, it has been shown in Refs.~\refcite{Previous1} and 
\refcite{Previous2} 
that a circles-in-the-sky detection of specific spatial topology 
can be used to reduce the degeneracies in the 
density parameter plane $\Omega_m-\,\Omega_{\Lambda}$, 
which arise from statistical analyses with data from current 
observations. 

Our main aim here, which are complementary to our previous 
works,\cite{Previous1,Previous2} is to show how one can use 
a possible circles-in-the-sky detection of the spatial topology 
of \emph{globally homogeneous} universes to set constraints 
on the dark energy equation of state (EOS) parameters.

\section{Topological Constraints and Concluding Remarks}
\label{MainRes}

To investigate how a possible detection of a nontrivial spatial topology 
can be used to place constraints on the dark energy equation of state 
parameters, we shall focus on  the globally homogeneous spherical manifolds 
and indicate how a similar procedure can be used in the case of globally
homogeneous flat topologies.\footnote{Since there are no Clifford 
translations in the hyperbolic geometry, there are no globally 
homogeneous hyperbolic manifolds.}

\begin{figure*}[htb!]
\begin{center}
\includegraphics[scale=0.4]{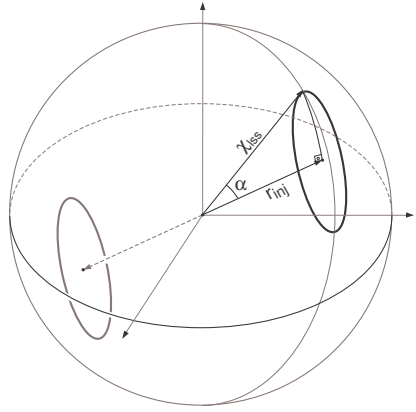} 
\caption{A schematic illustration of two antipodal
matching circles in the sphere of last scattering. These pairs of circles 
occur in all globally homogeneous universes with a detectable nontrivial 
topology. \label{Fig1}}
\end{center} \vspace{-4mm}
\end{figure*}

The topological constraints on dark energy EOS can be looked upon as
having two main ingredients, namely one of observational nature, and  
another of theoretical character.
Regarding the former, an important point about the globally homogeneous 
universes is that the pairs of topologically correlated circles on the
LSS will be antipodal, as shown in Figure~\ref{Fig1}.
A straightforward use of trigonometric relations for the right-angled 
spherical triangle shown in Fig.~\ref{Fig1} yields 
\begin{equation} \label{Obs_ChiLSS}
\cos \alpha = \frac{\tan r_{inj}}{\tan \chi^{}_{lss} } 
\qquad \mbox{or} \qquad
\chi^{}_{lss} = \tan^{-1} \left[\,\frac{\tan r_{inj}}{
\cos \alpha}\, \right] \,,
\end{equation}
where $r_{inj}$ is a topological invariant, whose values are given
in Table~\ref{Globally-Hom-Spherical}, and the distance $\chi^{}_{lss}$ 
is the comoving distance to the LSS \emph{in units of the present-day curvature 
radius}, $a_0=a(t_0)=(\,H_0\sqrt{|1-\Omega_{\mathrm{tot}}|}\,)^{-1}\,$ for
$k \neq 0$.

\begin{table}[ph] \vspace{-2mm}
\tbl{Globally homogeneous spherical manifolds along with their names, 
the covering groups and their order, and the injectivity radius
$r_{inj}$. For details on this notation and a brief description of 
the classification we refer the readers to Ref.~6 
and the appendix of Ref.~7. 
}
{\begin{tabular}{@{}cccc@{}} \toprule
 Manifold  & Covering Group $\Gamma$ & Order of $\Gamma$ & Injectivity Radius: $r_{inj}$  \\
\colrule
$\mathcal{Z}_n:=\mathbb{S}^3/Z_n$    &   Cyclic            $Z_n$   & $n$  & $\pi/n$  \\
$\mathcal{D}^*_m:=\mathbb{S}^3/D^*_m$& Binary dihedral     $D^*_m$ & $4m$ & $\pi / 2m$ \\
$\mathcal{T}:=\mathbb{S}^3/T^*$      & Binary tetrahedral  $T^*$   & 24   & $ \pi/6$  \\
$\mathcal{O}:=\mathbb{S}^3/O^*$      & Binary octahedral   $O^*$   & 48   & $\pi/8$  \\
$\mathcal{D}:=\mathbb{S}^3/I^*$      & Binary icosahedral  $I^*$   & 120   & $\pi/10$  \\
\botrule
\end{tabular} \label{Globally-Hom-Spherical} }
\end{table}

Now, a circles-in-the-sky detection of a given topology would give a value 
for the radius $\alpha$ along with an observational uncertainty 
$\sigma_\alpha$. These observational data along with Eq.~(\ref{Obs_ChiLSS}) 
and the usual error propagation formula, give the observational distance 
$\chi_{lss}^{}$ to the LSS and the associated uncertainty $\sigma_{lss}^{}\,$. 

The second important ingredient of the above mentioned topological constraints 
is related to the theoretical dark energy model. Indeed, the comoving distance 
to the last scattering surface in units of the curvature radius is given 
by
\begin{equation}
\label{ChiLSS}
\chi_{lss}^{\rm th} = \frac{d^{}_{lss}}{a_0} = \sqrt{|\Omega_k|}\,
\int_1^{1+z_{lss}} \, \frac{H_0}{H(x)} \,\, dx \,,
\end{equation}
where $d^{}_{lss}$ is the radius of the LSS, 
$x=1+z$ is an integration variable, $H$ is the Hubble 
parameter, $\Omega_k =1-\Omega_{\mathrm{tot}}$ is the curvature density 
parameter, and $z_{lss}=1089$.\cite{WMAP-Spergel:2003} Clearly, different
parametrizations of the equation of state $\omega_x=p_x / \rho_x$ give
rise to different Friedmann equations, i.e.,  different ratios $H(z)/H_0\,$. 
Thus, for example, assuming that the current matter content of the Universe 
is well approximated by a dust of density $\rho_m$ (baryonic plus dark matter)
along with a dark energy perfect fluid component of density $\rho_x$ 
and pressure $p_x$, for the parametrizations $\omega_x= \mathrm{const.} =\omega_0$,
$\omega_x= \omega_0 + \omega_1 z$ (Refs.~\refcite{HutterTurner2001} and
\refcite{WellerAlbrecht200}), and $\omega_x= \omega_0 + \omega_1\, z /(1+z)$ (Refs.~\refcite{Chevallier2001} and \refcite{Linder2003})
the Friedmann equation takes, respectively, the following forms:
\begin{eqnarray} 
\left({ \frac{H}{H_0} }\right)^2 \!\! &=& \Omega_{m0}(1+z)^{3}+\Omega_{k0}(1+z)^{2} + 
 \Omega_{x0}(1+z)^{3(1+\omega_0)} \,,\label{XCDM} \\
\left({ \frac{H}{H_0} }\right)^2&=& \Omega_{m0}(1+z)^{3} + \Omega_{k0}(1+z)^{2} 
+\Omega_{x0}(1+z)^{3(\omega_0-\omega_1 + 1)} \exp (3\omega_1 z) \,, \label{Linear} \\
\left({ \frac{H}{H_0} }\right)^2 &=& \Omega_{m0}(1+z)^{3}+\Omega_{k0}(1+z)^{2} 
+\Omega_{x0}(1+z)^{3(\omega_0+\omega_1 + 1)} \exp (- \frac {3\omega_1 z}{1+z}) 
\label{CPL} \,.
\end{eqnarray} 

In order to show how one can use the topology to set constraints on the
dark energy equation of state parameters we combine the two above-mentioned
ingredients by comparing the observational topological value $\chi_{lss}^{}$
for a given topology  with the theoretical values $\chi_{lss}^{\rm th}$ for 
any dark energy equation of state parametrization
[assuming a Gaussian distribution with mean given by Eq.~(\ref{ChiLSS})].
Thus, the constraints from a detectable spatial topology are 
taken into account in a $\chi^2$ statistical  analysis of any parameter
plane (as, for example, $\omega_0-\,\Omega_k\,$, $\omega_1-\,\Omega_k\,$ $\omega_0-\,\omega_1$)  by adding  a new term of the form 
\begin{equation} \label{chisqtop}
\chi^2_{\rm top}= \left(\frac{\chi_{lss}^{} -\chi_{lss}^{\rm th}}%
{\sigma_{\chi_{lss}}}\right)^2 
\end{equation}
to the remaining $\chi^2$ terms that account for other 
observational data sets. A concrete application of this result can
be found in Ref.~\refcite{SLR}, where by assuming the Poincar\'{e} 
dodecahedral space (PDS) as the circles-in-the-sky observable spatial 
topology, the current constraints on the equation of state parameters 
for the parametrizations~Eq.(\ref{XCDM}) and Eq.(\ref{CPL}) have been 
reanalyzed by using Type Ia supernovae data from the Legacy sample\cite{Astier2006} 
along with the baryon acoustic oscillations (BAO) peak in the large-scale correlation function of the Sloan Digital Sky Survey (SDSS),\cite{Eisenstein2005} and CMB shift
parameter,\cite{Bond1997} with and without the topological statistical term. 
It is shown that the PDS topology provides relevant additional constraints on 
the dark energy EOS parameters for the two-parameter Chevallier-Polarski-Linder parametrization\cite{Chevallier2001,Linder2003} [Eq.(\ref{CPL})], but negligible 
further constraints on $\omega_0$ of Eq.(\ref{XCDM}). The authors also show that a suitable 
Gaussian prior on the curvature density parameter $\Omega_k$ can mimic the 
role of the topology in such statistical analyses. 

Regarding the flat manifolds, we first note that there are three classes of  
multiply-connected manifolds of the form $\mathbb{R}^3/\,\Gamma$ which are globally
homogeneous, namely the $3$-torus class (compact in three directions), the class 
of chimney spaces (compact in two directions) and the slap space class (compact 
in one direction).
Second, Eq.~(\ref{Obs_ChiLSS}) clearly reduces to
\begin{equation} \label{Obs_ChiLSS_flat}
\chi^{}_{lss} = \,\frac{r_{inj}}{\cos \alpha}\,  \,,
\end{equation}
but now  since the curvature radius of Euclidean $3$-space is infinite, one 
cannot identify $a_0 = a(t=t_0)$ with the curvature radius. Therefore, in this 
case there is no natural unit of length, and one has to use, for example, 
megaparsecs (Mpc) as unit of length. Hence, the comoving distance to 
the LSS (for $c=1)$ is given by     
\begin{equation} \label{d_LSS}
\chi_{lss}^{\rm th} = d^{}_{lss} = \,
\int_1^{1+z_{lss}} \frac{1}{H(x)} \,\, dx \,.
\end{equation}

Finally, we note that Euclidean manifolds are not \emph{rigid}: even though 
topologically equivalent the manifolds of a specific class (fixed group $\Gamma$) 
of quotient flat manifolds may have different size.\footnote{Although 
diffeomorphic they are not necessarily globally isometric. For example, a
manifold in the $3$-torus class can be constructed by taking a 
parallelepiped (or particularly a cube) of any size and identifying 
the opposite faces by translations.} In this way, the injectivity 
radius $r_{inj}$ for a class of flat multiply-connected manifolds, is not a 
topological invariant (constant). Thus, one should estimate $r_{inj}$ 
on physical grounds, as for example by fitting the CMB 
data.\cite{Aurich-et-al20008}

To conclude, we emphasize that the above procedure may be employed 
for any specific globally homogeneous detectable topology and dark 
energy EOS parametrizations along with an arbitrary combination of 
data sets.

\vspace{-2mm}

\section*{Acknowledgments}

This work is supported by Conselho Nacional de Desenvolvimento 
Cient\'{\i}fico e Tecnol\'{o}gico (CNPq) - Brasil, under grant 
No. 472436/2007-4. M. J. Rebou\c{c}as thanks CNPq for the grant 
under which this work was carried out.  I am also grateful to 
A. Bernui for indicating misprints and omissions.

\end{document}